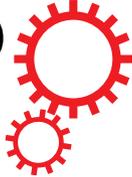

OPEN

# High-order exceptional points in optomechanics

H. Jing[1], Ş. K. Özdemir[2], H. Lü[3] & Franco Nori[4,5]

We study mechanical cooling in systems of coupled passive (lossy) and active (with gain) optical resonators. We find that for a driving laser which is red-detuned with respect to the cavity frequency, the supermode structure of the system is radically changed, featuring the emergence of genuine high-order exceptional points. This in turn leads to giant enhancement of both the mechanical damping and the spring stiffness, facilitating low-power mechanical cooling in the vicinity of gain-loss balance. This opens up new avenues of steering micromechanical devices with exceptional points beyond the lowest-order two.

Recent advances in cavity optomechanics (COM) have led to promising ways not only to observe the transition from classical to quantum worlds, but also to fabricate exotic devices for a wide range of applications, such as coherent wavelength conversion, optomechanically-induced transparency, phonon lasing, high-precision sensing, nonlinear acoustic control, quantum squeezing or chaos[1–4]. COM-based nonreciprocal optical control has also been achieved very recently[5,6]. The basic COM mechanism is the radiation-pressure-induced coherent light-sound coupling, which enables energy transfer either from sound to light (i.e., phonon cooling)[7–9] or from light to sound (i.e., phonon lasing)[10]. Despite considerable progress, it is still a highly challenging work to cool low-power COM devices deep into the quantum realm.

A promising new way of COM control is to utilize exotic properties of exceptional points (EP)[11–27], a form of degeneracy occurring in systems effectively described by non-Hermitian Hamiltonians (see refs [28–31]). Remarkable EP-assisted COM effects, e.g. low-power phonon emissions[28], chaos[29], and non-reciprocal energy transfer[31] or asymmetric mode switching based on dynamical EP-encircling[32], have been revealed. These studies, however, use a purely optical EP of order 2, where only two eigenfunctions of the system coalesce[31,33]. When more than two eigenfunctions coalesce[16–26], much richer physics has been revealed, e.g., enhanced spontaneous emission[23] or nanoscale sensing[27]. In a recent work, Teimourpour et al.[26] showed the creation of any number of high-order EP (h-EP) by using bosonic algebra. The realization and manipulations of h-EP have been proposed in purely optical systems, for examples, by using microdisks[19], waveguides[22], or photonic crystals[23]. We also note that the first experimental demonstration of h-EP has also been reported in coupled acoustic resonators[24]. However, up to date, the possible creation, the manipulations and the unconventional effects of h-EP have not been explored in hybrid COM devices.

In this work we study the evolutions of EPs in a three-mode COM structure, consisting of an active cavity coupled to a passive cavity supporting a mechanical mode[34–39]. We demonstrate that for an optical red detuning larger or smaller than the mechanical frequency $\omega_m$, only the EP of order 2 exists[28–30]; for the resonant case (i.e., the optical red detuning matches with the mechanical resonance frequency $\omega_m$), the supermode structure of the system is radically changed, featuring genuine high-order EP. As a result, in the vicinity of h-EPs, both the mechanical damping and the optical spring are significantly enhanced, facilitating low-power phonon cooling, with improved rates beyond what is achievable in conventional COM. These findings provide new insights for COM engineering with the aid of h-EPs and can be potentially useful for achieving various functional low-power acoustic devices.

## Results

We consider a system of two coupled microresonators, one with an optical gain $\kappa$ and the other with passive loss $\gamma$ [see Fig. 1(a)], such that both the coupling strength $J$ and the gain-to-loss ratio $\kappa/\gamma$ of the resonators can be tuned, as experimentally demonstrated in ref. 36. The mechanical mode (frequency $\omega_m$ and effective mass $m$), contained in the passive resonator, can be driven by an external field, with frequency $\omega_L$ and input power $P_{in}$[10]. The simplest Hamiltonian of this system can be written as ($\hbar = 1$).

[1]Key Laboratory of Low-Dimensional Quantum Structures and Quantum Control of Ministry of Education, Department of Physics and Synergetic Innovation Center for Quantum Effects and Applications, Hunan Normal University, Changsha, 410081, China. [2]Electrical and Systems Engineering, Washington University, St. Louis, Missouri, 63130, USA. [3]Key Laboratory for Quantum Optics, Shanghai Institute of Optics and Fine Mechanics, Chinese Academy of Science, Shanghai, 201800, China. [4]CEMS, RIKEN, Saitama, 351-0198, Japan. [5]Physics Department, University of Michigan, Ann Arbor, MI 48109-1040, USA. Correspondence and requests for materials should be addressed to H.J. (email: jinghui73@gmail.com) or Ş.K.Ö. (email: ozdemir@wustl.edu)

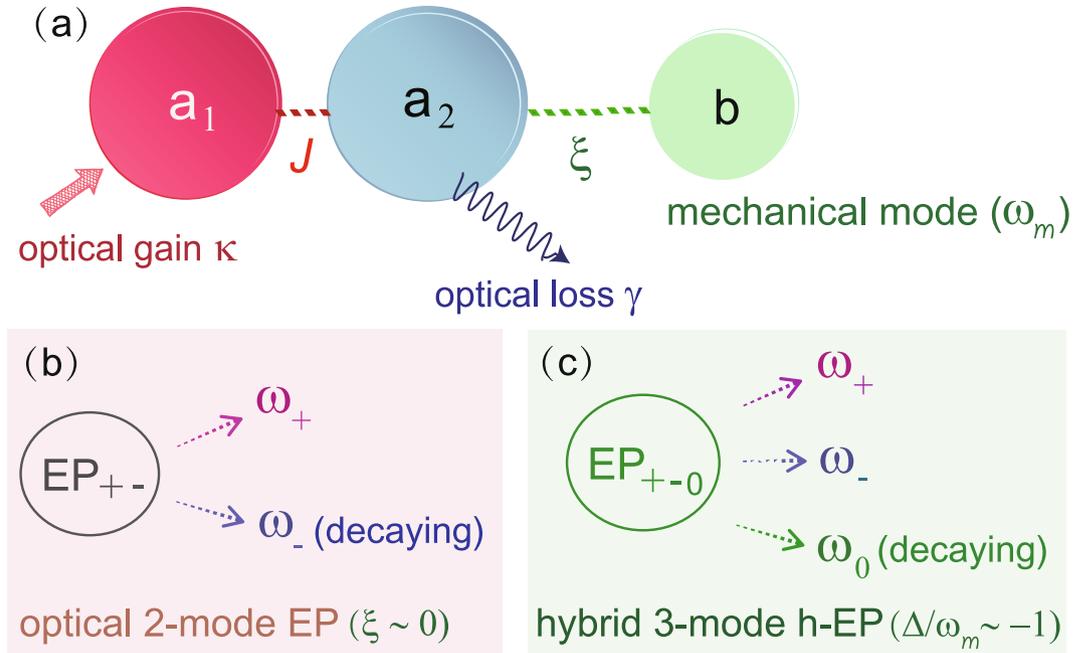

**Figure 1.** (a) Schematic illustration of a COM system with an active resonator coupled to a passive resonator containing a mechanical mode. (b) The off-resonance case has an EP which is of order 2, due to the coalescing optical modes[36]. This features a pair of amplifying and decaying optical supermodes, $a_\pm = (a_1 \pm a_2)/\sqrt{2}$, with frequencies $\omega_\pm$ (for details on $\omega_{0,\pm}$, see the Method). (c) When the optical red detuning equals $\omega_m$, an EP of order 3 emerges, leading to low-power phonon cooling in the vicinity of gain-loss balance (see the text).

$$\begin{aligned}
H &= H_0 + H_{int},\\
H_0 &= \frac{p^2}{2m} + \frac{1}{2}m\omega_m^2 x^2 - \Delta(a_1^\dagger a_1 + a_2^\dagger a_2),\\
H_{int} &= J(a_1^\dagger a_2 + \text{h.c.}) - \xi a_2^\dagger a_2 x + \eta_L(a_2^\dagger + a_2),
\end{aligned} \quad (1)$$

where $x$ and $p$ are the mechanical position and momentum operators, respectively, $J$ is the optical tunneling rate, $a_1$ and $a_2$ are the lowering operators for the optical modes in the resonators, $\xi = \omega_c/R$ is the COM coupling coefficient, with $R$ and $\omega_c$ respectively denoting the radius and the resonance frequency of the resonator supporting the mechanical mode, and $\eta_L = \sqrt{2\gamma P_{in}/(\hbar\omega_c)}$ is the pump rate. For phonon cooling, we choose a laser whose frequency is red-detuned with respect to the resonator, $\Delta = \omega_L - \omega_c < 0$.

The resulting equations of motion read

$$\begin{aligned}
\dot{a}_1 &= \kappa a_1 - iJa_2 + i\Delta a_1,\\
\dot{a}_2 &= -\gamma a_2 - iJa_1 + i\xi a_2 x + i\Delta a_2 - i\eta_L,\\
\dot{x} &= p/m,\\
\dot{p} &= -m\omega_m^2 + \xi a_2^\dagger a_2 - \Gamma_m p + \varepsilon_{th},
\end{aligned} \quad (2)$$

where $\gamma$ and $\Gamma_m$ are the optical and mechanical damping rates, respectively, and $\varepsilon_{th}$ denotes the thermal force at finite environmental temperature $T$, with zero mean value and the Brownian noise correlation

$$\langle \delta\varepsilon_{th}(t)\delta\varepsilon_{th}(t')\rangle = \frac{\Gamma_m}{2\omega_m}\int \frac{d\omega}{2\pi}\omega e^{i\omega(t'-t)}\left[1 + \coth\left(\frac{\hbar\omega}{2k_B T}\right)\right].$$

The steady-state solutions then become

$$x_s = \frac{\hbar |a_{2,s}|^2}{m\omega_m^2}, \quad p_s = 0, \quad a_{1,s} = \frac{iJa_{2,s}}{\kappa + i\Delta},$$

$$a_{2,s} = \frac{\eta_L(-\Delta + i\kappa)}{(J^2 - \kappa\gamma - \Delta^2 - \Delta\xi x_s) + i(\Delta\kappa - \Delta\gamma + \xi\kappa x_s)}.$$

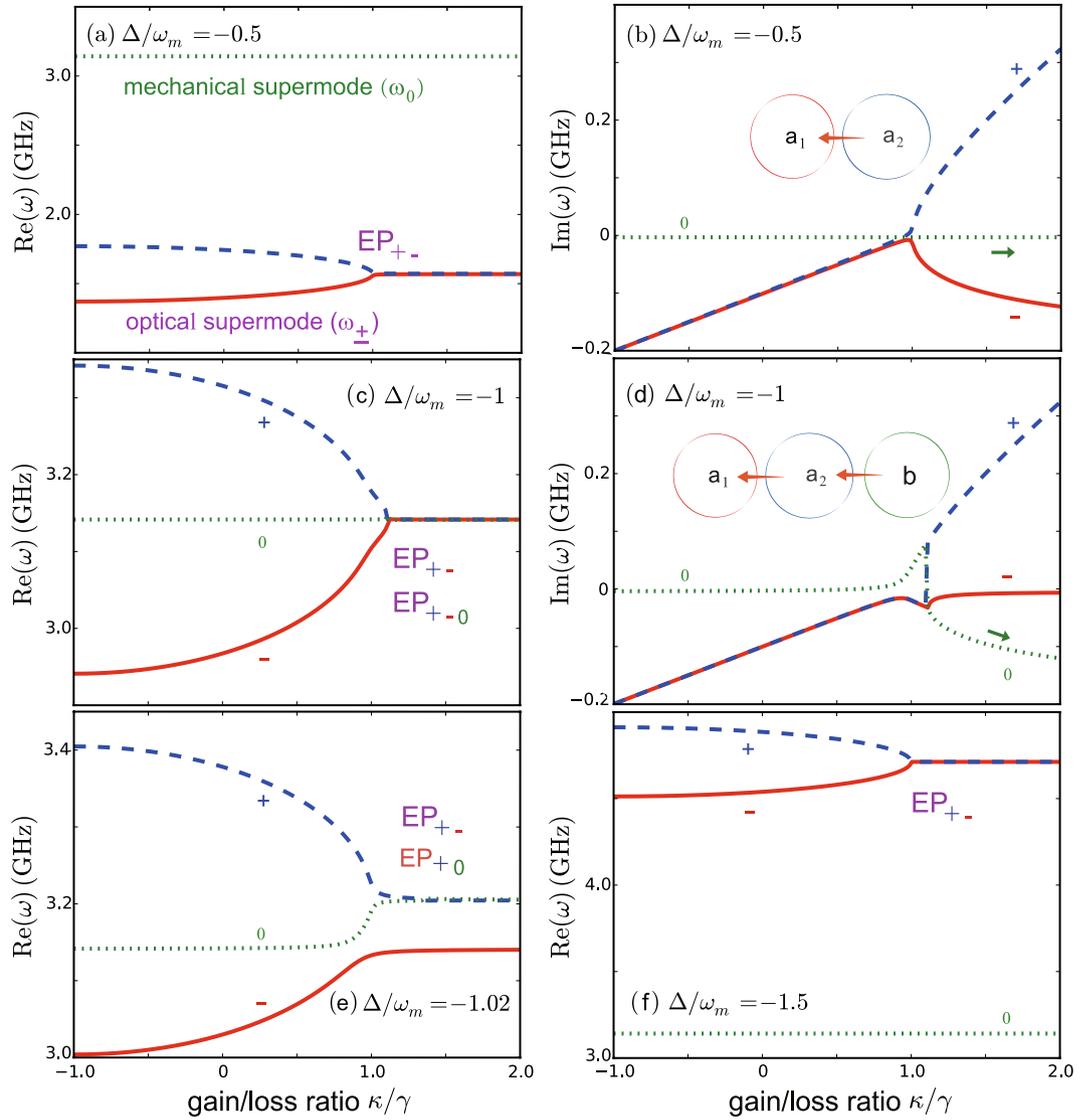

**Figure 2.** The supermode spectrum for the COM system in active-passive-coupled resonators ($J/\gamma = 1$). Here the input power is fixed as $P_{in} = 1$ mW.

Eq. (1) can be further linearized as that describing non-cyclic coupled three oscillators, for $\Delta < 0$[40], resulting in

$$H_{int} \rightarrow J(a_1^\dagger a_2 + \text{h.c.}) - (Ga_2^\dagger b + G^* b^\dagger a_2), \quad (3)$$

where

$$G = a_{2,s}\xi x_0, \quad x_0 = \sqrt{\hbar/2m\omega_m},$$

and $b$, the phonon operator, is defined by $x = x_0(b + b^\dagger)$. This linear three-mode model, including the optical gain and loss, can be exactly diagonalized (for more details, see the Method). The resulting supermode spectrum ($\omega_{\pm,0}$) is shown in Fig. 2, with the experimentally achievable values, i.e. $\lambda = c/\omega_c = 1550$ nm, $Q_c \sim 10^6$, $R \sim 20\,\mu$m, $\omega_m \sim 2\pi \times 500$ MHz, $m \sim 10$ pg, $Q_m \sim 10^3$, and $\xi = \omega_c/R \sim 10$ GHz/nm, $\gamma \sim 200$ MHz, $\Gamma_m \sim 3.14$ MHz[36, 39]. This type of spectrum, by tuning both $\kappa/\gamma$ and $\Delta/\omega_m$, is unattainable in passive COM[41].

Figure 2(a,b,f) show that, for optical red-detunings larger or smaller than the mechanical frequency $\omega_m$, the lowest two-order EP is observed for the supermodes with frequencies $\omega_\pm$, which is strongly reminiscent of what has been observed in all-optical devices[36] (hence we refer to $\omega_\pm$ as the optical-supermode frequencies), i.e., the imaginary parts of the eigenfrequencies bifurcate and lead to field localizations when surpassing the gain-loss balance. This feature, however, is radically changed near the COM resonance, where the optical red-detuning is equal to $\omega_m$. In this case an EP of order 3 emerges due to the interaction of the three modes of the system[16–26]. In particular, when surpassing the EP, energy transfer occurs among all the three supermodes [see Fig. 2(c,d)]. Similar

features are also seen for the near-resonance case, see Fig. 2(e) with $\Delta/\omega_m = -1.02$. This indicates a new way for mechanical cooling: using the interplay between $\kappa/\gamma$ and $\Delta/\omega_m$, which is not possible at all in conventional COM.

To demonstrate COM cooling at h-EPs (i.e., 3-EP in our system), we study the linear response of the system to external noise, by expanding all the operators around their mean values, $a_i \rightarrow a_{i,s} + \delta a_i$ ($i = 1, 2$). This yields the linearized equations in the matrix form

$$\dot{\vec{u}}(t) = A\vec{u}(t) + \vec{n}(t),  \qquad (4)$$

where $\vec{n}(t) = (0, 0, 0, 0, 0, \delta\varepsilon_{\text{th}})^T$ is the input noise vector, $\vec{u}(t) = (\delta X_1, \delta Y_1, \delta X_2, \delta Y_2, \delta x, \delta p)^T$ is the vector representing the noise of the dynamical variables of the system, with the redefined fluctuations

$$\delta X_i = (\delta a_i + \delta a_i^\dagger)/\sqrt{2}, \quad \delta Y_i = (\delta a_i - \delta a_i^\dagger)/i\sqrt{2},$$

and $A$ is the transfer matrix

$$A = \begin{bmatrix} \kappa & i\Delta & 0 & -iJ & 0 & 0 \\ i\Delta & \kappa & -iJ & 0 & 0 & 0 \\ 0 & -iJ & -\gamma & i\bar{\Delta} & 0 & 0 \\ -iJ & 0 & i\bar{\Delta} & -\gamma & 2i\xi a_{2,s} & 0 \\ 0 & 0 & 0 & 0 & 0 & \dfrac{1}{m} \\ 0 & 0 & R_2 & -iI_2 & -m\omega_m^2 & -\Gamma_m \end{bmatrix}, \qquad (5)$$

with effective detuning $\bar{\Delta} \equiv \Delta + \xi x_s$, and $(R_2, I_2) \equiv \sqrt{2}\hbar\xi(\text{Re}[a_{2,s}], \text{Im}[a_{2,s}])$.

We note that the rotating-wave approximation is invoked only in Eq. (4), not in Eq. (5). In addition, we have chosen the parameter values as above to satisfy the Routh-Hurwitz criteria of stability (i.e., the eigenvalues of $A$ have a non-positive real part), leading to stable solutions for Eq. (5)[28, 30]. The fluctuations of the pump field are important only when the pump light is modulated near $\omega_m$, where a back-action force larger than $\varepsilon_{\text{th}}$ appears, e.g. in the process of optomechanically-induced transparency[42–44]. For thermal-force-driven system, the response of the system to the thermal force $\varepsilon_{\text{th}}$ can then be obtained by solving Eq. (5) in the frequency domain, leading to

$$\delta x[\omega] = \lambda[\omega]\delta\varepsilon_{\text{th}}[\omega],  \qquad (6)$$

with the Lorentzian-type mechanical susceptibility

$$\lambda^{-1}[\omega] = m[(\Omega_{\text{eff}}^2 - \omega^2) - i\omega\Gamma_{\text{eff}}].  \qquad (7)$$

The effective mechanical frequency $\Omega_{\text{eff}}$ and damping rate $\Gamma_{\text{eff}}$ are then given by

$$\Omega_{\text{eff}}^2 = \omega_m^2 + \frac{\hbar\xi^2|a_{2,s}|^2}{m}\left\{\frac{(\bar{\Delta}+\omega)[\kappa(\kappa-\gamma) - A_-]}{A_-^2 + (\bar{\Delta}+\omega)^2(\kappa-\gamma)^2} + \frac{(\bar{\Delta}-\omega)[\kappa(\kappa-\gamma) - A_+]}{A_+^2 + (\bar{\Delta}-\omega)^2(\kappa-\gamma)^2}\right\},  \qquad (8)$$

$$\Gamma_{\text{eff}} = \Gamma_m - \frac{\hbar\xi^2|a_{2,s}|^2}{m}\left\{\frac{\kappa A_- + (\bar{\Delta}+\omega)^2(\kappa-\gamma)}{A_-^2 + (\bar{\Delta}+\omega)^2(\kappa-\gamma)^2} + \frac{-\kappa A_+ - (\bar{\Delta}-\omega)^2(\kappa-\gamma)}{A_+^2 + (\bar{\Delta}-\omega)^2(\kappa-\gamma)^2}\right\},  \qquad (9)$$

where,

$$A_\pm = (J^2 - \kappa\gamma - \bar{\Delta}^2 - \omega^2) \pm 2\Delta\omega.  \qquad (10)$$

Figure 3 shows the exact results for $\Omega_{\text{eff}}, \Gamma_{\text{eff}}$, by numerically solving Eqs (5–11). For comparison, the well-known results for a single lossy cavity are also shown in Fig. 3(a,b). We see that the conventional COM cooling is dominated by the optical damping, while the optical spring effect is negligible[45]. This situation is similar to that of the passive-passive resonators[39], as shown in Fig. 3(c,d). In contrast, Fig. 3(e,f) shows that for the compound system, by approaching $\kappa/\gamma = 1$, the optical spring effect becomes dominant, e.g. for $P_{\text{in}} \sim 0.1$ mW and $\kappa/\gamma \sim 1$, $\Omega_{\text{eff}}$ of the compound system is at least 3 orders of magnitude larger than that of the conventional COM system. The sharp sign inversion for $\Gamma_{\text{eff}}$, in the vicinity of $\kappa/\gamma = 1$ (Fig. 3f), means that when $\kappa/\gamma < 1$, the system operates in the amplified regime for the mechanical mode [see also Fig. 2(d)]; in contrast, when $\kappa/\gamma > 1$, it enters into the lossy regime (i.e., cooling of the mechanical mode). In other words, by surpassing the EP, the gain-to-loss compensation (for $\kappa/\gamma < 1$) evolves into field localizations (for $\kappa/\gamma > 1$), see also ref. 36. Clearly, approaching the exact balance of gain and loss (i.e., $\kappa/\gamma = 1$) from the right (i.e., $\kappa/\gamma \gtrsim 1$) results in efficient COM cooling due to the enhancement of both the optical spring and the effective mechanical damping.

In sharp contrast with conventional COM, the active-passive COM system features the low-power strong optical spring effect, in the vicinity of the gain-loss balance. This consequently results in a significant decrease in the phonon number

$$n = \frac{k_B T}{\hbar\omega_m}\frac{\Gamma_m}{\Gamma_{\text{eff}}}\left(\frac{\omega_m}{\Omega_{\text{eff}}}\right)^3 \propto \Gamma_{\text{eff}}^{-1}\Omega_{\text{eff}}^{-3}.$$

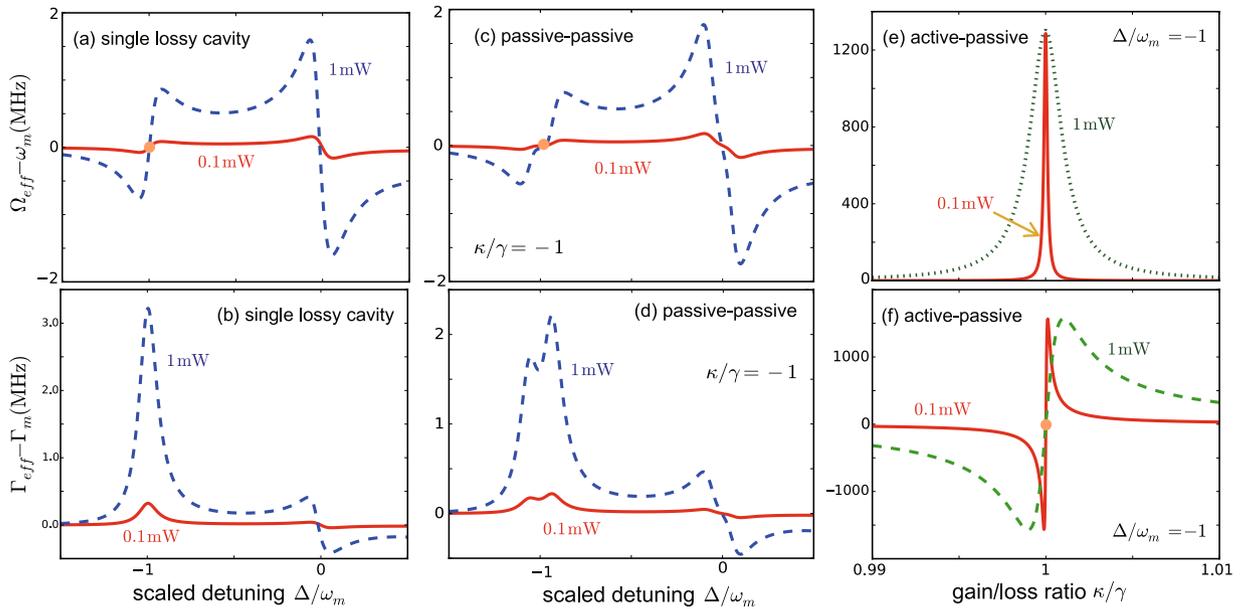

**Figure 3.** The effective mechanical frequency and damping rate for (**a,b**) the single passive resonator, (**c,d**) the passive-passive resonators, or (**e,f**) the active-passive resonators, with the fixed value $J/\gamma = 1$.

For comparisons, we have calculated the phonon number $n_0$ in conventional COM with a lossy cavity (see the Method). When $P_{in} = 1$ mW, $\Delta/\omega_m = -1$ (red detuning), we find that, by starting from room temperature, $n_0$ can be decreased down to $\sim 10^{3\,45}$ (see also the Method). A single active cavity, having the same COM parameters, is confirmed to exhibit similar cooling rates for blue detuning $\Delta/\omega_m = 1$ (see also ref. 46). In contrast, for the compound system, the tunable parameter $\kappa/\gamma$ provides a new way to enhance the cooling rate, without requiring blue detuning or strong COM coupling. Defining $n$ as the phonon number in the active-passive compound system, we use

$$\beta = n/n_0 \qquad (11)$$

to evaluate the cooling efficiencies of COM systems with passive-passive cavities (i.e., $\kappa/\gamma < 0$) and with active-passive cavities (i.e., $\kappa/\gamma > 0$). When $\kappa/\gamma < 1$, cooling at the sidebands is only slightly enhanced, see Fig. 4(a). This situation, as aforementioned, is radically changed when surpassing the h-EP [see also Fig. 3(d)]. For example, the factor $\beta$ is decreased by 3 orders of magnitude in the vicinity of the balance ($\kappa/\gamma \sim 1$) for $P_{in} = 0.1$ mW [see Fig. 4(b)]. This cooling rate at 0.1 mW is two orders of magnitude higher than that of a single lossy cavity driven at $P_{in} = 1$ mW, implying a significant benefit in terms of power budget. We have also calculated $\beta$ for other values of $P_{in}$, and found a reversed-dependence on $P_{in}$ (at $\sim 0.12$ mW), which is a general feature observed when a system is brought to its EPs (see also ref. 12). Nevertheless, for $P_{in} = 1$ mW, more than 2-order enhancement of the cooling can still be achieved, see Fig. 4(c). The ground-state cooling, or $n \sim 0.01$, is accessible with $P_{in}$ as low as 0.1 mW, when the system is first cooled down to 650 mK [Fig. 4(d)]. This is two orders of magnitude improvement in the cooling rate compared to ref. 8 which reported $n \sim 1.7$ with a power of 1.4 mW when the system temperature was 650 mK. Our approach utilizing an EP of order 3 also allows achieving $n \lesssim 1$ with a power of 0.12 mW at temperatures as high as 20 K (Fig. 4(d), see also the Method). This opens up the prospect to engineer low-power micromechanical devices based on exotic high-order EPs.

## Discussion

In conclusion, we have studied the emergence and applications of high-order EPs in a COM device composed of coupled active and passive cavities[36]. We find that for a pump laser that is red-detuned from the cavity resonance by the mechanical frequency (i.e., $\Delta/\omega_m \sim -1$), the supermode structure of the system radically changes, featuring emergence of h-EP. We stress that in such a non-Hermitian COM system, the high-order EPs (i.e., the coalesce of three super-modes) appear only when the optical red detuning matches with the mechanical resonance frequency; otherwise, it is reduced to the lowest EP of order-2, i.e., the coalesce of two optical-like super-modes, which is similar to a purely optical system. This new possibility of observing the tunable high-order EPs in COM has not been reported previously. In addition, we find that in the vicinity of high-order EPs, significant enhancement of both the optical spring and the mechanical damping leads to a new route to achieve low-power mechanical cooling. Our findings open up novel prospects for applications of h-EP in realizing low-power COM or quantum acoustic devices. Future efforts along this direction include the study of e.g., h-EP in nonlinear COM[47–50] or phononic engineering by dynamically encircling the EPs[31, 32].

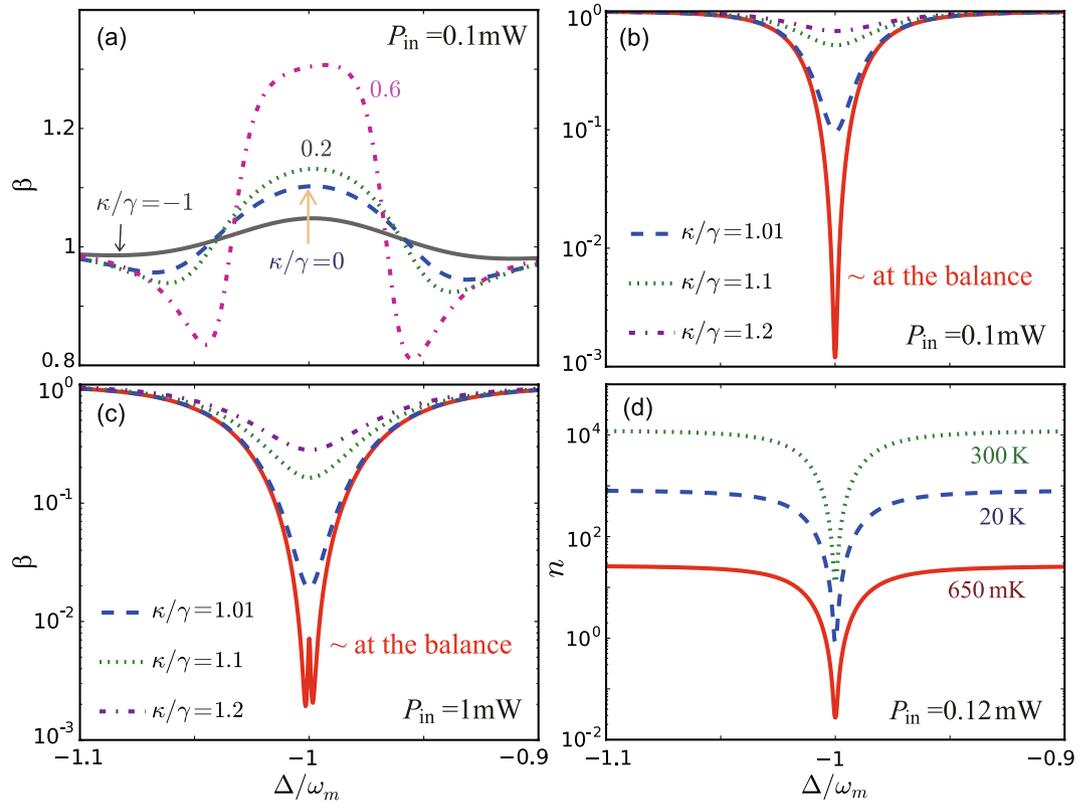

**Figure 4.** (**a**–**c**) The cooling enhancement factor $\beta = n/n_0$, as a function of $\Delta/\omega_m$ for various values of $\kappa/\gamma$, when (**a**,**b**) $P_{in} = 0.1$ mW, and (**c**) $P_{in} = 1$ mW. (**d**) Achievable minimum phonon number $n$ at the gain-loss balance for $P_{in} = 0.12$ mW, when the initial temperature $T$ is 300 K, 20 K, or 650 mK (i.e., cryogenic temperature[8]), respectively.

## Methods

**More details about the supermode structure.** By linearizing the COM coupling $\xi a_2^\dagger a_2 x$ in the optical red-detuning regime, the non-Hermitian Hamiltonian of the system, including the optical gain and loss, can be written at the simplest level as ($\hbar = 1$)

$$H_{lin} = (-\Delta - i\kappa)a_1^\dagger a_1 + (-\Delta + i\gamma)a_2^\dagger a_2 + \omega_m b^\dagger b + J(a_1^\dagger a_2 + a_2^\dagger a_1) - (Ga_2^\dagger b + G^* b^\dagger a_2), \quad (12)$$

where $G = a_{2,s}\xi x_0$, $x_0 = (2m\omega_m)^{-1/2}$, $\Delta = \omega_L - \omega_c$, and $b$ is the annihilation operator of the mechanical mode, with $x = x_0(b + b^\dagger)$. The optical weak driving terms are not explicitly shown here. This linearized three-mode Hamiltonian, with noncyclic inter-mode interactions, can be exactly diagonalized in the super-mode picture[51–53]. The resulting super-mode frequency $\omega$ is determined by the determinant equation

$$\begin{vmatrix} \omega - (-\Delta - i\kappa) & J & 0 \\ J & \omega - (-\Delta + i\gamma) & -G \\ 0 & -G^* & \omega - \omega_m \end{vmatrix} = 0, \quad (13)$$

or, more explicitly, the cubic equation

$$\omega^3 + \lambda_1 \omega^2 + \lambda_2 \omega + \lambda_3 = 0, \quad (14)$$

with

$$\begin{aligned}
\lambda_1 &= 2\Delta + i(\kappa - \gamma) - \omega_m, \\
\lambda_2 &= [-2\Delta + i(\gamma - \kappa)]\omega_m + \Delta^2 + i(\kappa - \gamma)\Delta + \kappa\gamma - |G|^2 - J^2, \\
\lambda_3 &= [J^2 - \Delta^2 - i(\kappa - \gamma)\Delta - \kappa\gamma]\omega_m + |G|^2(-\Delta - i\kappa).
\end{aligned} \quad (15)$$

Clearly, for $\xi = 0$, we have

$$\begin{vmatrix} \omega - (-\Delta - i\kappa) & J & 0 \\ J & \omega - (-\Delta + i\gamma) & 0 \\ 0 & 0 & \omega - \omega_m \end{vmatrix} = 0, \tag{16}$$

and the familiar results in an all-optical two-mode system can be obtained[36], where only an EP of the lowest-order 2 exists for the optical supermodes $a_{\pm} = (a_1 \pm a_2)/\sqrt{2}$. The difference between the resonance frequencies and between the linewidths of the supermodes in the strong and weak coupling regimes are, respectively, given by

$$\Delta\omega \equiv \omega_+ - \omega_- = 2\left[J^2 - \left(\frac{\kappa+\gamma}{2}\right)^2\right]^{1/2}, \quad \Delta\gamma \equiv \gamma_+ - \gamma_- = 0, \quad \text{for} \quad \kappa < 2J - \gamma,$$

$$\Delta\omega = 0, \quad \Delta\gamma = 2\left[\left(\frac{\kappa+\gamma}{2}\right)^2 - J^2\right]^{1/2}, \quad \text{for} \quad \kappa > 2J - \gamma. \tag{17}$$

As shown in Fig. 2(a,b,f), a similar supermode-splitting feature also appears for the far-off-resonance case, with $\xi \neq 0$. For an analytical confirmation, we take as a specific example $J/\gamma \sim 1$ and $\Delta/\omega_m \sim -0.5$. To solve the cubic equation Eq. (14), we introduce $w = x - \frac{\lambda_1}{3}$, so that Eq. (14) can be written as

$$x^3 + px + q = 0, \tag{18}$$

where

$$\begin{aligned} p &= \lambda_2 - \frac{1}{3}\lambda_1^2 \\ &= -\frac{1}{3}(\Delta + \omega_m)^2 + \frac{1}{3}(\kappa - \gamma)(\kappa + 2\gamma) - \frac{i}{3}(\kappa - \gamma)(\Delta + \omega_m) - |G|^2, \\ q &= \lambda_3 - \frac{1}{3}\lambda_1\lambda_2 + \frac{2}{27}\lambda_1^3 \\ &= -\frac{2}{27}(\Delta + \omega_m)^3 - \frac{i}{9}(\kappa - \gamma)^2(\Delta + \omega_m) - \frac{i}{9}(\kappa - \gamma)(\Delta + \omega_m)^2 \\ &\quad - \frac{i}{27}(\kappa - \gamma)^2(2\kappa + 7\gamma) + \frac{|G|^2}{3}[i(\kappa - \gamma) - (\Delta + \omega_m)]. \end{aligned} \tag{19}$$

By using Cardano's formula[54], we have the solutions for Eq. (18)

$$\begin{aligned} x_1 &= u + v, \\ x_2 &= \left(-\frac{1}{2} + i\frac{\sqrt{3}}{2}\right)u + \left(-\frac{1}{2} - i\frac{\sqrt{3}}{2}\right)v, \\ x_3 &= \left(-\frac{1}{2} - i\frac{\sqrt{3}}{2}\right)u + \left(-\frac{1}{2} + i\frac{\sqrt{3}}{2}\right)v, \end{aligned} \tag{20}$$

where

$$\begin{aligned} u &= \left(-\frac{q}{2} + \sqrt{\frac{p^3}{27} + \frac{q^2}{4}}\right)^{\frac{1}{3}} \approx u_0 + \frac{|G|^2}{2(\Delta + \omega_m)^2}[i(\kappa - \gamma) - (\Delta + \omega_m)], \\ v &= \left(-\frac{q}{2} - \sqrt{\frac{p^3}{27} + \frac{q^2}{4}}\right)^{\frac{1}{3}} \approx v_0 - \frac{|G|^2}{2(\Delta + \omega_m)^2}[i(\kappa - \gamma) - (\Delta + \omega_m)], \end{aligned} \tag{21}$$

with

$$\begin{aligned} u_0 &= \frac{\Delta + \omega_m}{3} + \sqrt{\gamma^2 - \left(\frac{\kappa + \gamma}{2}\right)^2} - i\frac{\kappa - \gamma}{6}, \\ v_0 &= -\frac{\Delta + \omega_m}{3} - \sqrt{\gamma^2 - \left(\frac{\kappa + \gamma}{2}\right)^2} - i\frac{\kappa - \gamma}{6}. \end{aligned} \tag{22}$$

Then we have

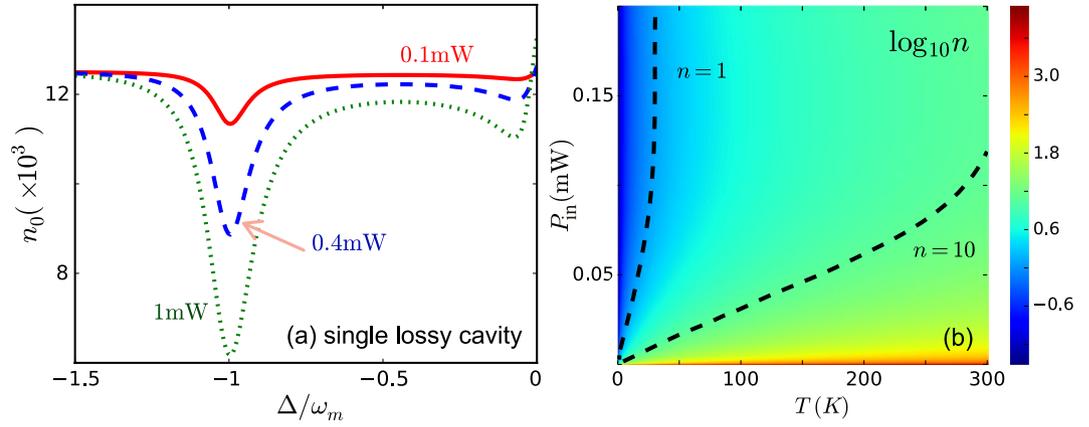

**Figure 5.** (**a**) The phonon number $n_0$ for the conventional COM composed of a single passive resonator at different powers when the system is initially at room temperature. The parameter values used in this simulation is the same as those given in the main text. (**b**) The phonon number for the COM system composed of active-passive-coupled resonators, with respect to the temperature and the input pump power.

$$\begin{aligned}
\omega_0 &\approx \omega_m, \\
\omega_+ &\approx \Delta + \omega_m + \sqrt{\gamma^2 - \left(\frac{\kappa+\gamma}{2}\right)^2} - \frac{\sqrt{3}\,|G|^2(\kappa-\gamma)}{2(\Delta+\omega_m)^2} + i\frac{\kappa-\gamma}{2}, \\
\omega_- &\approx \Delta + \omega_m - \sqrt{\gamma^2 - \left(\frac{\kappa+\gamma}{2}\right)^2} + \frac{\sqrt{3}\,|G|^2(\kappa-\gamma)}{2(\Delta+\omega_m)^2} + i\frac{\kappa-\gamma}{2}, \text{ for } \kappa < 2J - \gamma,
\end{aligned}$$
(23)

and

$$\begin{aligned}
\omega_0 &\approx \omega_m, \\
\omega_+ &\approx \Delta + \omega_m + i\left(\sqrt{\gamma^2 - \left(\frac{\kappa+\gamma}{2}\right)^2} + \frac{\kappa-\gamma}{2} + \frac{\sqrt{3}\,|G|^2}{2(\Delta+\omega_m)}\right), \\
\omega_- &\approx \Delta + \omega_m - i\left(\sqrt{\gamma^2 - \left(\frac{\kappa+\gamma}{2}\right)^2} - \frac{\kappa-\gamma}{2} - \frac{\sqrt{3}\,|G|^2}{2(\Delta+\omega_m)}\right), \text{ for } \kappa > 2J - \gamma,
\end{aligned}$$
(24)

This indicates that

$$\begin{aligned}
\Delta\omega &\equiv \sqrt{4\gamma^2 - (\kappa+\gamma)^2} - \frac{\sqrt{3}\,|G|^2(\kappa-\gamma)}{(\Delta+\omega_m)^2},\ \Delta\gamma \equiv \gamma_+ - \gamma_- \simeq 0, \text{ for } \kappa < 2J - \gamma, \\
\Delta\omega &\simeq 0,\ \Delta\gamma = \sqrt{4\gamma^2 - (\kappa+\gamma)^2} + \frac{\sqrt{3}\,|G|^2}{\Delta+\omega_m}, \text{ for } \kappa > 2J - \gamma.
\end{aligned}$$
(25)

Except for an almost unchanged supermode (with frequency $\omega_0$), the familiar two-order EP can be observed for the supermodes $a_\pm$, which is strongly reminiscent of what has been observed in all-optical devices[36]. Hence, we refer to $a_\pm$ as the optical supermodes (i.e., the imaginary parts of the eigenfrequencies bifurcate and lead to field localization when surpassing the EP) [see also Fig. 2(a,b) of the main text]. We have confirmed that these analytical results agree well with the full numerical results as shown in Fig. 2 of the main text. The exact analytical solutions of the cubic equation, including the complicated results for the case with $\Delta/\omega_m \sim -1$ and the cumbersome relations between $(a_\pm, a_0)$ and $(a_{1,2}, b)$[51–53], are irrelevant to our work here. For the full numerical results of the supermode spectrum, see Fig. 2 of the main text.

**Phonon number in a single passive resonator.** For comparison purposes, we give here the minimum attainable phonon number $n_0$ in the conventional COM composed of a lossy cavity [Fig. 5(a)] using the same system parameters given in the main text. We find that $n_0$ can be decreased down to $\sim 10^3$ for $P_{in} = 1$ mW, when the system is initially at room temperature $T = 300$ K. This agrees well with the experiment reported in ref. 45. As shown in the main text, the cooling rate for the active-passive-coupled system driven by a lower power $P_{in} = 0.1$ mW is two orders of magnitude higher than the cooling rate achieved for a lossy cavity (conventional COM) driven by a higher power $P_{in} = 1$ mW. This implies a significant benefit in terms of power budget [see also Fig. 5(b)].

### Acknowledgements
H.J. is supported by NSFC Grant No. 11474087. S.K.O. is supported by ARO Grant No. W911NF-16-1-0339. F.N. is supported by the RIKEN iTHES Project, the MURI Center for Dynamic Magneto-Optics via the AFOSR award number FA9550-14-1-0040, the IMPACT program of JST, CREST, and a Grant-in-Aid for Scientific Research (A).

### Author Contributions
H.J. conceived the idea and performed the calculations with the help from H.L.; H.J. and S.K.Ö. wrote the manuscript with input from F.N.; all the authors discussed the content of the manuscript.

### Additional Information
**Competing Interests:** The authors declare that they have no competing interests.

**Publisher's note:** Springer Nature remains neutral with regard to jurisdictional claims in published maps and institutional affiliations.